\documentclass[aip, apl, pbalance, amsmath,amssymb, lengthcheck,reprint]{revtex4-2}
\usepackage{xcolor}
\usepackage{graphicx}
\usepackage{enumitem}
\usepackage{mathptmx,mathtools}
\usepackage{balance}
\usepackage[cmbraces,varbb,varvw]{newtxmath}
\usepackage{etoolbox}
\usepackage[scaled=1.0]{helvet}
\usepackage{titlesec}
\usepackage{soul}
\usepackage[bookmarks = false, colorlinks = true, linkcolor = black, citecolor = black, urlcolor = black, linkbordercolor = white, pdfstartview={XYZ null null 1.00}]{hyperref}
\usepackage{fancyhdr}
\titleformat{\section}
  {\normalfont\sffamily\large\bfseries\color{black}}
  {\thesection.}{1em}{}
\titleformat{\subsection}
  {\normalfont\sffamily\large\bfseries\color{black}}
  {\Alph{subsection}.}{1em}{}

\usepackage{filecontents}

\makeatother
\begin{document}

\preprint{AIP/123-QED}

\title[\textit{\footnotesize Applied Physics Letters}]{Modulating outcomes of oil drops bursting at a water-air interface}
\author{Varun Kulkarni}
\altaffiliation[Corresponding author, electronic mail: ]{\href{varun14kul@gmail.com}{varun14kul@gmail.com}}
\affiliation{Department of Mechanical and Industrial Engineering, University of Illinois at Chicago, Chicago, IL 60607, USA \looseness=-1}
\author{Suhas Tamvada}
\affiliation{Department of Mechanical and Industrial Engineering, University of Illinois at Chicago, Chicago, IL 60607, USA \looseness=-1} 
\author{Yashasvi Venkata Lolla}
\affiliation{Department of Mechanical and Industrial Engineering, University of Illinois at Chicago, Chicago, IL 60607, USA \looseness=-1}
\author{Sushant Anand}
\altaffiliation[Corresponding author, electronic mail: ]{\href{sushant@uic.edu}{sushant@uic.edu}}
\affiliation{Department of Mechanical and Industrial Engineering, University of Illinois at Chicago, Chicago, IL 60607, USA \looseness=-1}

\begin{abstract}
Recent studies have shown that capillary waves generated by bursting of an oil drop at the water-air interface produces a daughter droplet inside the bath while part of it floats above it. Successive bursting events produce next generations of daughter droplets, gradually diminishing in size until the entire volume of oil rests atop the water-air interface. In this work, we demonstrate two different ways to modulate this process by modifying the constitution of the drop. Firstly, we introduce hydrophilic clay particles inside the parent oil drop and show that it arrests the cascade of daughter droplet generation preventing it from floating over the water-air interface. Secondly, we show that bursting behavior can be modified by a compound water-oil-air interface made of a film of oil with finite thickness and design a regime map which displays each of these outcomes. We underpin both of these demonstrations by theoretical arguments providing criteria to predict outcomes resulting therein. Lastly, all our scenarios have a direct relation to control of oil-water separation and stability of emulsified solutions in a wide variety of applications which include drug delivery, enhanced oil recovery, oil spills and food processing where a dispersed oil phase tries to separate from a continuous phase.
\\ Area: Interdisciplinary Applied Physics, Surfaces and Interfaces.
\end{abstract}

\maketitle


Dispersion of oil droplets in aqueous solutions are common in household items like salad dressings \citep{Nikolova2023}, ointments\citep{Takamura1984} and cosmetic creams\citep{Tadros1992} but even at larger scales in environment and industry such as oil spills\citep{Kulkarni2021,Kulkarni2018} and enhanced oil recovery\citep{Mandal2010}. These interactions are characterized by either their ability to remain together as a homogeneous mixture or get separated as disparate oil and water phases. In this light, two major directions for research have emerged over the years, on the one hand, the focus has been on understanding the kinetics and enhancing stability of dispersed oil droplets in a continuous water phase\citep{Borwankar1992, Guha2017, Liu2015} while on the other, mechanisms for their separation have been investigated. Results from these findings have often found use in applications like emulsion synthesis\citep{Wu2018} and dispersal of oil slicks modulated by surfactants\citep{Kuang2024}, microbes\citep{Ghosh2021}, particles\citep{Liu2019} or the combination thereof\citep{Li2022}. While the role of chemical constitution and its symbiotic relationship with physical hydrodynamics in separation or homogenizing of oil-water mixtures using physio-chemical modifiers like surfactants or particles is known\citep{Ngai2014, Anand2024}, it is yet unclear whether purely hydrodynamic mechanisms, which rely only on interfacial forces can be the main drivers of such two-phase interactions. 

Along this line, recent work by \textit{Kulkarni et al.}\citep{Kulkarni2024, Lolla2019} has revealed a novel, unexplored pathway of oil separation from the surrounding water phase by investigating the bursting of a rising oil drop at an air-water interface. It was demonstrated that the bursting oil drop produces a daughter droplet within the continuous phase and the process cascades down until the oil drop form a film above the water-air interface. The significance of this may be readily appreciated, as leakages from broken underwater oil pipelines and natural seeps release oil droplet plumes where individual oil drops rise \citep{Wang2023}, eventually bursting at the water-air interface and may even be encountered in two-phase microfluidic droplet-based flows. Quite naturally, methods to control this separation provide interesting avenues for future investigations, with profound implications. 

In this work, we pursue this objective and demonstrate two specific methods to control and modify this behavior by manipulating the oil drop/water interface which heretofore has only been shown by changing the viscosity of the water bath\citep{Kulkarni2024, Lolla2019}(continuous phase). In the first, we show that hydrophilic Bentonite (clay) particles mixed in an oil drop rising in a water bath can self-assemble at the oil-water interface to controllably halt successive bursting events. This situation resembles crude oil leaking from underwater pipeline bursts and entrains mineral/clay particles from the seafloor\citep{Payne2003, Hill2002, Khelifa2002} or that seen in separation of Pickering emulsions\citep{Zheng2022}. Our second study is inspired by double emulsion synthesis and buoyant crude oil jets entrapping the surrounding water phase to form a water-in-oil compound drop\citep{Xue2019}. We consider bursting of such compound drops made of encapsulated water drops of varying diameters which heretofore been only investigated for oil-coated air bubbles \citep{Ji2022, Yang2023a, Yang2023b} or falling drops \citep{Deka2019}. By these simple modifications to the oil drop-water interface we may regulate bursting outcomes such as cessation of bursting cascade at any given stage corresponding to a particular daughter droplet size and morphology of the daughter droplet produced.
\begin{figure*}[t!]
   \vspace{0pt}
	\centerline{\includegraphics[width=\textwidth]{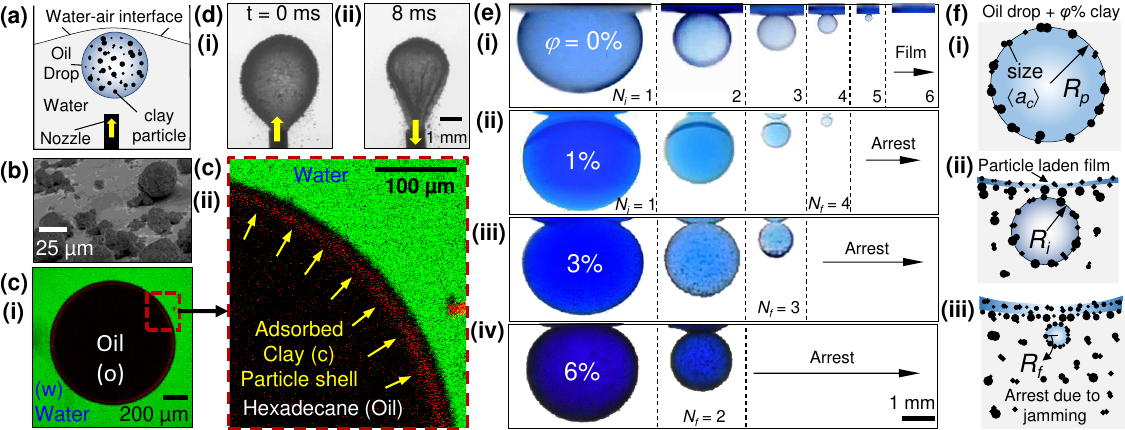}}
   \vspace{-5pt}
   \caption{(\textit{a}) Schematic of experimental setup showing a rising oil drop containing bentonite clay particles bursting at an air-water interface after being released upwards. (\textit{b})  Scanning electron microscopy (SEM) image of bentonite clay particles with a median size of 20 $\mu$m. (\textit{c})-(\textit{i}) Confocal laser scanning microscope (CLSM) image of an hexadecane (oil) drop in water with adsorbed clay particles at the oil-water interface forming a shell. Clay particles and bulk water are labeled using Rhodamine-B fluorescent red dye and Fluorescein green dye respectively with hexadecane not being dyed (therefore seen in black). (\textit{c}) -(\textit{ii}) Zoomed view of (\textit{c})-(\textit{i}) showing the adsorption of particles at the oil-water interface. Yellow arrows show the shell of clay particles (dyed red) (\textit{d}) Proof of self assembly of clay particles demonstrated by, (\textit{d})-(\textit{i}) injecting a particle-laden oil drop into a bulk of water  and, retracting the liquid (\textit{d})-(\textit{ii}). Self assembled particles adsorbed at the oil-water interface form a shell around the oil droplet which crumples during retraction. Effect on cascade(Multimedia available online), (\textit{e})(\textit{i}) without particles\citep{Kulkarni2024}, (\textit{ii})-(\textit{iv}) with particles showing early cessation with increasing particle concentration, $\varphi =$ 1, 3 and 6 \% (\textit{f}) Mechanism of cascade arrest, (\textit{i}) Initial particle coverage (\textit{ii}) Drop bursting leading to particle-laden oil film and daughter droplet generation with increased surface coverage below the interface (\textit{iii}) Formation of a \textit{Pickering} drop with final arrest with 90\% drop surface coverage of particles.}
   \label{Fig1}
   \vspace{-15pt}
\end{figure*}

For our experiments\citep{Kulkarni2024, Lolla2019}, we use hexadecane ($\rho_o$, 773 kg$\cdot$m\textsuperscript{-3} and dynamic viscosity, $\mu_o$, 3$\times10^{-3}$ Pa$\cdot$s) oil drops, insoluble in a D.I. (deionised) water bath ($\rho_w$, 998 kg$\cdot$m\textsuperscript{-3} and dynamic viscosity, $\mu_w$, 1$\times10^{-3}$ Pa$\cdot$s). In addition to its practical relevance, since water or hexadecane do not significantly dampen the capillary waves as in the case of more viscous outer liquids or typical two-fluid drop dynamics \citep{Rahman2018a, Rahman2018b}, we choose them as the continuous (bulk) liquid and dispersed (drop) phase, respectively to observe all subsurface phenomena clearly. The interfacial tension, $\sigma_{w/o}$ between hexadecane and water is 0.052 N$\cdot$m\textsuperscript{-1} and $\mu_o$ of hexadecane is small enough for viscous effects to be neglected\citep{Kulkarni2021}. Furthermore, hexadecane has a spreading coefficient of -0.0083 N$\cdot$m\textsuperscript{-1} which makes it non-spreading, forming a lenticular shape above the water-air interface. Its spreading is also considered negligible since its time scale is much longer than that for the subsurface dynamics\citep{Suhas2020}. The experimental observations are recorded using videos\citep{Kulkarni2021} taken at 4500 fps and a resolution of 1024 $\times$ 1024 pixels such that 1 pixel $\approx$ 15 $\mu m$. The depth of the water bath is maintained such that when the drops are completely detached from the needle they continue to be completely submerged inside the water bath. Additional details of the experiments are presented below in the appropriate sections.

To investigate the effect of inclusion of particles on oil drop bursting at the water-air interface a known initial weight, $\varphi \approx 3\%$ of hydrophilic bentonite clay particles was introduced into the parent (\textit{p}) oil (hexadecane) drop as illustrated in Fig. \ref{Fig1}\textcolor{black}{(\textit{a})}. The diameter of the clay (\textit{c}) particles measured using SEM varied between 5$-$25 $\mu$\textit{m} with an average value, $\langle a_{c} \rangle$ of 15 $\mu$\textit{m}, where, $\langle \cdot \rangle$ stands for the average of particles of different sizes (see Fig. \ref{Fig1}\textcolor{black}{(\textit{b})}). The particle-laden oil drop is released from a 0.5 mm inner diameter nozzle producing a parent drop of radius, $R_p$ equal to 2.2 mm and volume of nearly 45 $\mu$\textit{l}. The bentonite clay particles are hydrophilic in nature \citep{Abend2001} and almost completed wetted by water due to which the oil-water interfacial tension (and surface energy) is not affected much and we only use them in weight percentages between 0-6\% which does not alter the overall oil drop density, significantly. Lastly, these particles have low solubility in water but their negative surface charge is neutralized by it, depressing their zeta potential \citep{Huang2016} and not considered as a dominant factor in our analysis

As the oil drop rises through water due to the hydrophilic nature of the clay particles they are expected to adsorb\citep{Yadav2019} and self-assemble at the oil-water interface. To confirm this we conducted two specific experiments. In our first study, we used confocal laser scanning microscopy (CLSM) with water labeled using a green fluorophore and clay particles using a red one. The self-assembly of the clay particles upon release of particle-laden oil drop to form a shell (red ring) encapsulating the drop surrounded by water is clearly visualized here and shown in \ref{Fig1}\textcolor{black}{(\textit{c})(\textit{i})} (taken mid-plane, cutting across the diameter of the drop). The enlarged view in \ref{Fig1}\textcolor{black}{(\textit{c})(\textit{ii})} shows details of part of the drop revealing the formation of a 20 $\mu$\textit{m} thick particle shell covering the oil drop of radius 1 mm. 

To further confirm these observations we conducted a second set of experiments in which we dispensed an oil drop containing clay particles from a nozzle allowing the hydrophilic particles to self-assemble at the oil-water interface and form a shell as depicted in Fig. \ref{Fig1}\textcolor{black}{(\textit{d})(\textit{i})}, $t = 0$ \textit{ms}. Thereafter, we slowly retract oil from the nozzle and notice discernible creasing/crumpling of the interface after a few moments, at $t = 8$ \textit{ms} (see Fig. \ref{Fig1}\textcolor{black}{(\textit{d})(\textit{ii})}), owing to the presence of shell of clay particles. 
\begin{figure}[t!]
   \vspace{0pt}
	\includegraphics[width=\columnwidth]{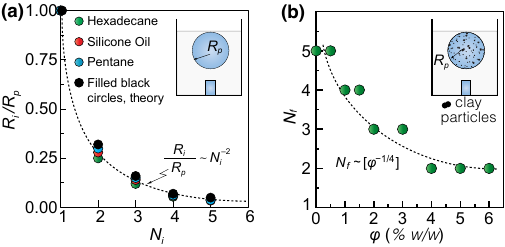}
   \vspace{-18pt}
   \caption{(\textit{a}) Ratio of daughter droplet $R_{i}$ and parent drop radius, $R_p$ at each bursting event, $N_i$ in the absence of particles in three oils (hexadecane, silicone oil, pentane) \citep{Kulkarni2024} drop yielding the scaling, $R_{f}/R_p \sim N_i^{-2}$. (\textit{b}) Decrease in number of bursting events required arrest cascade $N_f$ with increasing clay particle concentration ($\varphi$) in the oil drop exhibiting a scaling dependence of the form, $N_f \sim [\varphi^{-1/4}] $. Refer Fig. \ref{Fig1}\textcolor{black}{(\textit{e})} and \textcolor{black}{(\textit{f})} for symbols used here.}
   \label{Fig2}
   \vspace{-18pt}
\end{figure}
These above results confirm self-assembly of solid layer of clay particles around the oil drop enabling us to explore the effect of increasing $\varphi$ on the bursting of hexadecane (oil) drops. To this end, we varied $\varphi$ from 1 to 6 wt\% and recorded the number of bursting events, $N_f$ required to arrest the daughter droplet cascade. The sequence of images obtained from videos (Video 1, Multimedia available online) recorded from our experiments is shown in Fig. \ref{Fig1}\textcolor{black}{(\textit{e})(\textit{i})-(\textit{iv})}. The first row represents the case when particles were absent in the oil drop and it is observed that after $N_i = 5$ the entire original parent oil drop formed a film (without clay particles) as reported in our previous work \cite{Kulkarni2024}. On introducing clay particles at increasing $\varphi$ of 1\%, 3\% and 6\%, as shown in the second, third and the fourth row, the cascade is arrested much earlier, at $N_f = 4, 3$ and 2 respectively (Video 1, Multimedia available online) producing a particle covered drop, reminiscent of a \textit{Pickering} drop of the anti-Bancroft type \citep{Binks2006, Golemanov2006, Zheng2022} each time, which was stable for atleast 18 hours until the water bath completely evaporated.

In order to understand the underlying mechanism, we consider the particle-laden parent oil drop of radius $R_p$ after it is released from the nozzle as portrayed schematically in Fig. \ref{Fig1}\textcolor{black}{(\textit{f})(\textit{i})} and denoted by the symbol, $N_i = 1$, where the subscript, \textit{i} denotes the generation of daughter droplet with $i =1$ being the parent drop. 

During the rise of the parent oil drop, clay particles preferentially self-assemble\citep{Binks2006, Ngai2014} along the water-oil interface with a certain initial surface coverage, $\chi_1$ ultimately halting near the water-air interface. This is followed by bursting of thin bulk (\textit{b}) water film between the oil drop and water-air interface producing a daughter droplet\citep{Kulkarni2024} (of radius, $R_{i}$) with higher intermediate particle surface coverage, $\chi_i$ compared to the initial particle-laden parent drop and a thin layer particle-infused oil film atop the water-air interface as sketched in Fig. \ref{Fig1}\textcolor{black}{(\textit{f})(\textit{ii})}. With subsequent bursting events ($N_i > 1$) the daughter droplet size, $R_{i}$ continues to decrease with increasing $\chi_i$ eventually resulting in tight enough packing of particles (with coverage $\chi_f$) on its drop/bulk interface at which stage ($N_f$) subsequent bursting is arrested. No more oil can drain through this particle shell at this stage and bursting is jammed by the particles-infused oil film residing above (see Fig. \ref{Fig1}\textcolor{black}{(\textit{f})(\textit{iii})}). Even though it might follow from (initial) high value of $\chi_1$ that bursting will not commence, such a configuration does not lead to stable oil-in-water emulsion, restricting $\varphi$ to 0 to 6 wt. \% (see SM, Sec. 1 for details and clay particle size distribution). \textcolor{black}{Using the expression by Golemanov \citep{Golemanov2006}, $\varphi = 8\chi_1(\rho_c/\rho_o)/ ((2R_p/\langle a_c\rangle) + 8\chi_1(\rho_c/\rho_o - 1))$ and plugging in values of $\chi_1 =$ 0.9, $\rho_c =$ 2400 kg$\cdot$m\textsuperscript{-3}, $\rho_o =$ 773 kg$\cdot$m\textsuperscript{-3}, $R_p =$ 2.2 mm, $\langle a_c\rangle =$ 15 $\mu$m for anti-bancroft emulsions we obtain, $\varphi \approx$ 7\% for which a drop of radius 2.2 mm will be stable by itself. We have chosen a limit slightly below this value of 6\% to ensure we continue to see atleast one bursting event.}
\begin{figure*}[t!]
   \vspace{0pt}
	\centerline{\includegraphics[width=\linewidth]{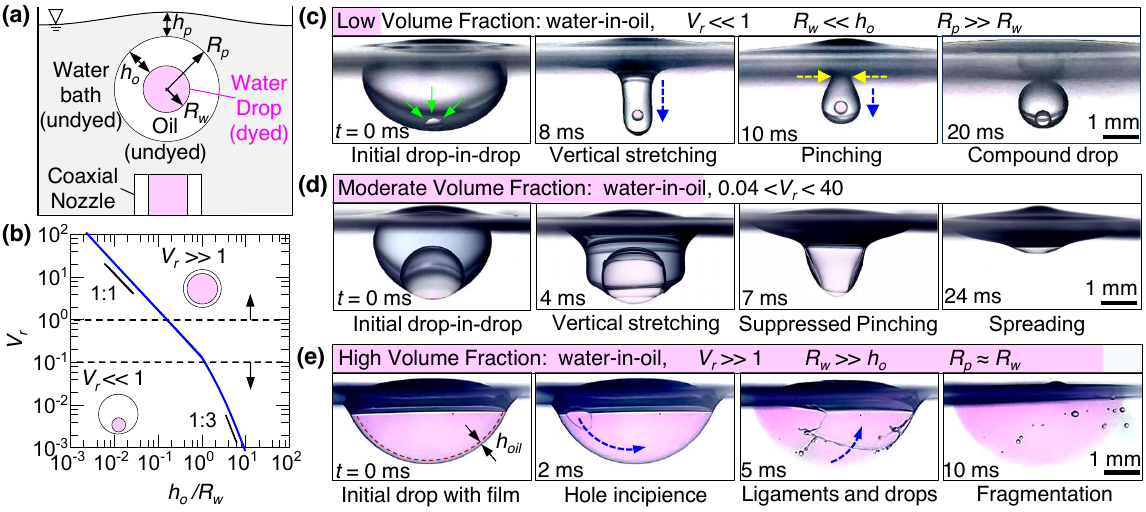}}
   \vspace{-10pt}
   \caption{(\textit{a}) Schematic shows a rising water-in-oil compound drop generated using a co-axial nozzle. The inner water drop is dyed pink using rhodamine D while the water bath and the oil covering it are not. (\textit{b}) Dependence of volume ratio, \textit{V\textsubscript{r}} of water-in-oil in a compound drop on the ratio of oil film thickness and water drop radius, \textit{h\textsubscript{o}/R\textsubscript{w}}. In the limit, $V_{r} << 1$ it reduces to $V_{r} \approx (h_{o}/R_w)^{1/3}$ and for $V_r >> 1$ it yields, $V_{r} \approx (3h_{o}/R_w)^{-1}$. Bursting process of a compound drop (Multimedia available online) (\textit{c}) at low $V_{r} \approx 5\%$ (green arrows, initial encapsulated water drop) producing an encapsulated daughter water drop of higher $V_{r}$ (dotted arrows, movement direction) (\textit{d}) at intermediate $0.04 < V_{r} < 40$ with suppression of daughter droplet production. (\textit{e}) at high $V_{r} \approx 97\%$, leading to underwater oil film fragmentation (dotted arrows, hole expansion direction) and polydispersed daughter oil droplets. Note, waves after bursting approximately travel a distance of $1.5\pi R_p$ to cover the entire drop. }
   \label{Fig3}
   \vspace{-15pt}
\end{figure*}

To make quantitative predictions about $N_f = F\left(\varphi\right)$ based on the above mechanism, we note that until the formation of a final stable daughter droplet of radius, $R_{f}$, the bursting process is expected to be similar to the case when particles are absent (see Fig. \ref{Fig1}\textcolor{black}{(\textit{e})(\textit{i})}) since $\chi_{i=1}$ is not large enough to cover the entire drop's surface initially. However, once $\chi_{i}$ reaches $\chi_{f} = 0.90$, the cascade is arrested\citep{Golemanov2006} by the mechanism explained previously. We could use this to relate $R_{f}$ with final number of particles, $n_{c,f}$ that it contains using the relation\citep{Golemanov2006}, $\chi_{f} = {n_{c,f}\langle a_{c} \rangle^2 }/{16 R_{f}^2}$. For a constant $\langle a_{c} \rangle$ and $\chi_{f} = 0.9$ (at arrest) we may write, $R_{f}^2 \sim n_{c,f}$. From here, it only remains to connect $n_{c,f}$ with $\varphi$ and $R_{f}$ with $N_f$ to get the functional form of $N_f$ in terms of $\varphi$. 

In pursuit of the above, we first recognize, $\varphi = \rho_{c} \langle a_{c} \rangle^3 n_{c,i=1}/8\rho_o R_p^3$ where, $n_{c,i=1}$ the initial number of clay particles and $\rho_{c} \approx 2400$ kg$\cdot$m\textsuperscript{-3} is density of clay particles\citep{Blake2008}. Since, $R_p \approx 3.4\;$mm, $\langle a_{c} \rangle$ and $\rho_o$ are constants for our experiments we obtain, $\varphi \sim n_{c,1}$. Moreover, larger $\varphi$ requires larger $R_{f}$ with a larger surface area, $\sim R_{f}^2$ (see previous paragraph) to accommodate all the particles, therefore, it is reasonable to expect, $\varphi \sim n_{c,1} \sim n_{c,f} \sim R_{f}^2$ thereby simplifying our scaling relation to the form, $R_{f}^2 \sim \varphi$.

Now, it only remains to connect the number of bursting events, $N_f$ required to arrest the cascade at a specific daughter droplet of radius, $R_{f}$. To do so, we use the scaling relation derived in our previous work \citep{Kulkarni2024} for the daughter droplet size,  $R_{i =2} \sim \zeta^{-2.34}R_p^{2.74}$ where the constant, $\zeta=\mu_b/\sqrt{\rho_b\sigma_{pb}}$. To obtain the daughter droplet size after the next bursting event, $N_i = 3$ we substitute $R_{i=2}$ from this scaling as $R_p$ in the same expression. Recursively, this gives rise to, $R_{i=3} \sim \zeta^{-2.34(1 + 2.74)} R_p^{2.74^2}$. Therefore, after $N_i \left(\geq 2\right)$ cascades we write the general expression for $R_{i}$, the daughter droplet size after $N_i$ generations, $R_{i}/R_p \sim \zeta^{-2.34 \mathcal{M}} R_p^{2.74^{N_i -1}-1}$, where, $\mathcal{M} = \sum_{i = 1}^{N_i -1} 2.74^{i-1}$. We use this expression to determine theoretically the variation of the daughter droplet size, $R_{i}/R_p$ with $N_i$ for a given $R_p \approx 3.4$ mm which we see to match our experimental values accurately (see Fig. \ref{Fig2}\textcolor{black}{(\textit{a})}). However, this form is inconvenient to use for further analysis and therefore we seek a more concise functional dependence of the type, $R_{i}/R_p = F\left(N_i\right)$. To do so, we use a best fit curve to represent our experimentally and theoretically obtained values and obtain the relation, $R_{i} \sim R_p N_i^{-2}$ which for a constant $R_p \approx 3.4\;$ mm reads, $R_{i} \sim N_i^{-2}$. This relation could alternatively even be obtained rigorously by expanding $\zeta^{-2.34 \mathcal{M}} R_p^{2.74^{N_i -1}-1}$ in Taylor series to obtain its polynomial form but not undertaken here for conciseness. At the arrest radius, $R_f$ this assumes the form, $R_{f} \sim N_f^{-2}$, obtained by setting $i =f$ in $R_{i} \sim N_i^{-2}$. Finally, we combine this scaling relation with $R_{f}^2 \sim \varphi$ which gives number of bursting events or cascades, $N_{i=f}$ before arrest at an initially prescribed $\varphi$ as,
\begin{equation}\label{eqn4}
N_f \sim [\varphi^{-1/4}]
\end{equation}
In  Eq. \ref{eqn4} $[\cdot]$, is the nearest integer function and the scaling obtained accurately predicts the data as shown in Fig. \ref{Fig2}\textcolor{black}{(\textit{b})} with a scaling prefactor evaluated as $1.2 \pm 0.03$. \textcolor{black}{We also state that the scaling relation (Eq. {\hypersetup{linkcolor=black}\ref{eqn4}}) remains unchanged for any particle size distribution (see Sec. S1 SM) or any initial parent drop radius, $R_p$ as it algebraically cancels out leading to Eq. ({\hypersetup{linkcolor=black}\ref{eqn4}}).}

Our results so far show that solid shell of particles provide an attractive route to modify post-bursting outcomes. Encouraged by these findings, we examine the consequence of bursting at an air-water interface when a liquid shell of variable thickness surrounds around the oil drop as rationalized by a water-in-oil-compound drop. To study this we use a coaxial nozzle of inner diameter 0.56 mm and annular gap of 1.7 mm nozzle producing a compound parent drop of radius, $R_p = $ 2.2 mm and volume $V_p$ with varying water (\textit{w}) drop radius $R_w$, (0.5-2.1 mm) and volume, $V_w$ producing an oil (\textit{o}) layer of thickness, $h_o$ and volume $V_o$ as shown schematically in Fig. \ref{Fig3}\textcolor{black}{(\textit{a})} which is subjects the bulk (water) and oil layer inside the drop to continual drainage (see Sec. S2 SM). Three volume fractions, $V_r:=V_w/V_o$ which correspond to $V_{r} <<1$, $V_{r} >> 1$ and $0.04 < V_{r} < 40$ of the compound drops are tested whose dependence on dimensionless oil layer thickness, $h_{o}/R_w$ is shown in Fig. \ref{Fig3}\textcolor{black}{(\textit{b})}. For these calculations we compute, $V_{r}$ as $R_{w}^{3}/(R_{p}^{3} - R_{w}^{3})$. Considering $R_p = R_w + h_o$ we can rewrite $V_r$ as, $[(h_{o} R_{p}^{2}/R_{w}^{3})(1+R_{w}/R_{p}+R_{w}^{2}/R_{p}^{2})]^{-1}$. In the limit, $V_r <<1$ and $V_r >>1$ this reduces to, $h_{o}/R_{w} \approx V_r^{-1/3}$ and $h_{o}/R_{w} \approx \frac{1}{3}V_r^{-1}$ respectively (more details of this algebra may be found in Sec. S3, SM)

Our experimental results at different $V_{r}$ are shown in Fig. \ref{Fig3}\textcolor{black}{(\textit{c})-(\textit{e})} (also see Video 2, Multimedia available online). At low, $V_{r} \left(<< 1\right)$ once the parent compound drop bursts, capillary waves descend downwards pinching the drop to form a daughter droplet which encapsulates the original water drop within (see Fig. \ref{Fig3}\textcolor{black}{(\textit{c})}). In emulsion synthesis where excess material removal has been typically achieved through solvent evaporation\citep{Sheth2020, Vian2018} this could provide a facile solution strategy. At intermediate $0.04 < V_{r} < 40$ the encapsulated water drop is large enough to suppress daughter droplet formation altogether as shown in Fig. \ref{Fig3}\textcolor{black}{(\textit{d})} and drop just floats up eventually. Finally, at large, $V_{r} \left(>> 1\right)$ the thin oil film surrounding the water drop bursts at the bulk water-air interface producing sub-surface polydispersed oil droplets (see Fig. \ref{Fig3}\textcolor{black}{(\textit{e})}) bearing similarities to bursting of curved thin liquid films \citep{Kulkarni2023, Kulkarni2014} and those produced by raindrop impacts on oil slicks\citep{Murphy2015}.

\begin{figure}[t!]
   \vspace{0pt}
	\centerline{\includegraphics[width=\columnwidth]{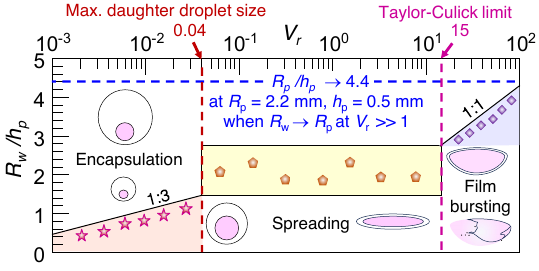}}
   \vspace{-11pt}
   \caption{Design map predicting encapsulation, spreading and film bursting at different dimensionless water drop radius, $R_w/h_p$ as a function of $V_r$ for $R_p = 2.2$ mm. Symbols are experimental data.}
   \label{Fig4}
   \vspace{-20pt}
\end{figure}

To mathematically determine when each of these regimes will be observed we develop a design map based on our experimental data as shown in Fig. \ref{Fig4}. Our arguments follow from the premise that rising water-in-oil compound drops experiences two competing drainage flows whose time scales are determined by Stefan-Reynolds theory\citep{Nguyen2000}(see Sec. S2, SM). At $V_r << 1$, the time scale of drainage, $\sim \mu_oR_w^4/F_{w/o}h_o^2$ of the oil layer ($h_o >> R_w$) due to apparent weight ( $F_{w/o}$) of the encapsulated water drop within the oil (parent, hexadecane) drop given by, $\left(\rho_w - \rho_{o}\right)V_wg + \rho_wV_pg$ (where, $g = 9.81$ m$\cdot$s\textsuperscript{-2} is gravitational acceleration) is required to be greater than the time scale, $\sim \mu_wR_p^4/F_{p/b}h_p^2$ of buoyancy driven squeezing of bulk water film of thickness $h_p$ driven by the force, $F_{p/b} = \left(\rho_w - \rho_{o}\right)V_pg$ to ensure water drop remains inside the compound daughter droplet after bursting. \textcolor{black}{This condition leads us to the balance, $(\rho_r -1)(2 \rho_{r} V_{r}-V_{r}+\rho_{r})^{-1} \sim \mu_{r} (h_{\text{o}}^{2}/R_{w}^{2})(R_{w}^{2}/h_{p}^{2})$ at the regime boundary, where, $\rho_r := \rho_w/\rho_o$ and $\mu_r := \mu_w/\mu_o$}. Using the geometric approximation, $h_{o}/R_w \approx V_{r}^{-1/3}$ (see Fig.\ref{Fig3}\textcolor{black}{(\textit{b})}) and algebraic simplification, \textcolor{black}{$2 \rho_{r} V_{r}-V_{r}+\rho_{r} \approx \rho_{r} $} for $V_r << 1$ (see Sec. S3, S4, SM) and considering that the fluid properties, $\rho_r = 1.2 $, $\mu_{r} = 3$ and $R_p = 2.2$ mm) are a constant we arrive at the following simplified relation for the boundary of encapsulation,
\begin{equation}\label{eqn5}
R_w/h_p \sim V_r^{1/3}
\end{equation}
Similarly, to derive the criterion of film bursting at $V_{r} >> 1$ the time scale of drainage of the \textit{thin} oil film ($h_o << R_w$) due to dominant interfacial tension or capillarity (\textit{cap}), $F_{\text{\textit{cap}}} = \sigma_{w/o}\left(2\pi R_p\right)$ should be greater than the buoyancy driven drainage due to force, $F_{p/b}$ described above \textcolor{black}{which at the regime boundary reads, $F_{p /\text{b}}/F_{\text{cap}}\sim \mu_{r} h_{\text{o}}^{2}/h_{p}^{2}$. This simplifies to, $(\rho_r-1)(2/3)[(\rho_r -1)(1+ V_r)]^{-1}(Bo_{w/\text{o}}) \sim \mu_{r} h_{\text{o}}^2/h_p^2$ where, $Bo_{w/\text{o}} = (\rho_w - \rho_o)R_p^2\mathrm{g}/\sigma_{w/o}$}. Like above, fluid properties, $\mu_r$, $\rho_r$ and $R_p$ are a constant which along with the geometric approximation, $h_{o}/R_w \approx (3V_{r})^{-1}$ (see Sec. S3, S4, SM and Fig. \ref{Fig3}\textcolor{black}{(\textit{b})}) yields at the transition boundary, 
\begin{equation}\label{eqn6}
R_w/h_p \sim V_{r}
\end{equation}
For closure, we evaluate three more bounds for $R_p \approx 2.2$ mm: (I) since lubrication flow approximation is valid for Reynolds number, $Re_b = \rho_w v_w h_p/\mu_w << 1$; for a rise velocity, $v_w \approx 0.002$ m/s (from experiments) we obtain $h_p = 0.5$ mm which results in, $R_p/h_p \approx 4.4$. \textcolor{black}{Note that maximum value attained by $R_w/h_p$ is when the encapsulated water drop occupies the entire volume of the compound drop, at $V_{r} >> 1$, equivalent to $R_w \to R_p$ and therefore, $R_w/h_p =  R_p/h_p \approx 4.4$.} (II) to ensure complete encapsulation after bursting we require, $R_w \leq R_{i=2}= R_p/2^2$ where, $R_2$ is the daughter droplet radius after first bursting event (see Fig. \ref{Fig2}). For, $h_{o}/R_w \approx V_{r}^{-1/3}$ at $V_r << 1$ and $h_o/R_w \approx R_p/R_w$ this results in, $V_r \approx$ 0.04 (III) Lastly, at $V_r >> 1$ the oil film transitions from capillary dominated dynamics at moderate $V_r$ at a time scale\citep{Deka2019}, $\tau_{mod} = (\rho_o R_p^3/\sigma_{w/o})^{0.5}$ with capillary waves traveling a distance $\approx 1.5\pi R_p$ at an average velocity scale, $v_{mod}/2 = 1.5\pi R_p/\tau_{mod}$ (see Fig. \ref{Fig3}\textcolor{black}{(e)}) to bursting of a curved film\citep{Lhuissier2012} at higher $V_r$ with an expanding hole retracting at the Taylor-Culick velocity\citep{Lhuissier2012}, $v_{tc} = (2\sigma_{w/o}/\rho_oh_o)^{0.5}$. For smooth crossover, $v_{tc}$ equals $v_{mod}$ which simplifies to, $h_o/R_p = (9\pi^2/2)^{-1}$. On using the geometric approximation $V_r \approx (3h_o/R_w)^{-1}$ for $V_r >> 1$ and $R_w \approx R_p$ we ultimately obtain the bound, $V_r \approx$ 15. The limits (I), (II) and (III) (see Sec. S4, SM for details) along with Eqs \ref{eqn5} and \ref{eqn6} shown in Fig. \ref{Fig4} complete the design map. 
\\
\indent In summary, we show two ways to tailor consequence of a bursting oil drop at an air-water interface. First, we introduce hydrophilic clay particles inside the oil drop to arrest daughter droplet generation and, second, we encapsulate a water droplet inside an oil drop to find distinct behaviors at various water to oil volume fractions. The former of our demonstrations is directly connected to practical situations like entrainment of clay particles from the ocean bed in underwater oil pipeline bursts and \textit{Pickering} emulsions. The latter is inspired by water encapsulation in buoyant oil jets, optimizing size of double emulsions and raindrop impact on oil slicks. \textcolor{black}{In practical scenarios as seen in oceans and seas, we expect humidity, temperature, surfactants and other contaminants to be present\citep{Zinke2022}. These can delay or accelerate the bursting processes but not prevent their observance altogether as reported in our work. Hence, we do expect our results to be relevant in normal marine climatic conditions and serve as guide for any future detailed investigations.} In addition to oceanic/atmospheric sciences and colloidal synthesis for drug delivery/food/cosmetics, applications that benefit from ingenious manipulation of the consequences of bursting drops could also find our results useful.
\vspace{-18pt}
\section*{\normalsize Acknowledgment}
\vspace{-10pt}
Financial support for this project through NSF (EAGER) award no. 2028571 is gratefully acknowledged. The authors thank Navid Saneie for his assistance in obtaining the SEM image for bentonite clay particles.
\vspace{-18pt}
\section*{\normalsize Supplementary Material (SM)}
\vspace{-10pt}
See the supplementary material (SM) accompanying this manuscript that contains (\textit{i}) consequences of high initial particle coverage in arresting daughter droplet formation and effect of particle size distribution (\textit{ii}) additional details of algebra leading to the scaling inequalities, (\ref{eqn5}) and (\ref{eqn6}) and bounds, I, II and III.
\vspace{-18pt}
\section*{\normalsize Data Availability Statement}
\vspace{-10pt}
Data underlying the conclusions of the paper are available in the plots presented and can be provided upon request.
\vspace{-18pt}
\section*{\normalsize Conflict of Interest Statement}
\vspace{-10pt}
The authors have no conflict of interests to disclose.
\vspace{-4pt}
\balance
\bibliography{APLRefs}

\end{document}


\maketitle
\tableofcontents

\clearpage

\section{Reasons for limiting particle weight \% to $\varphi = 6\%$}
\begin{figure}[htp!]
      \vspace{-10pt}
   \centering
    \includegraphics[width = \textwidth]{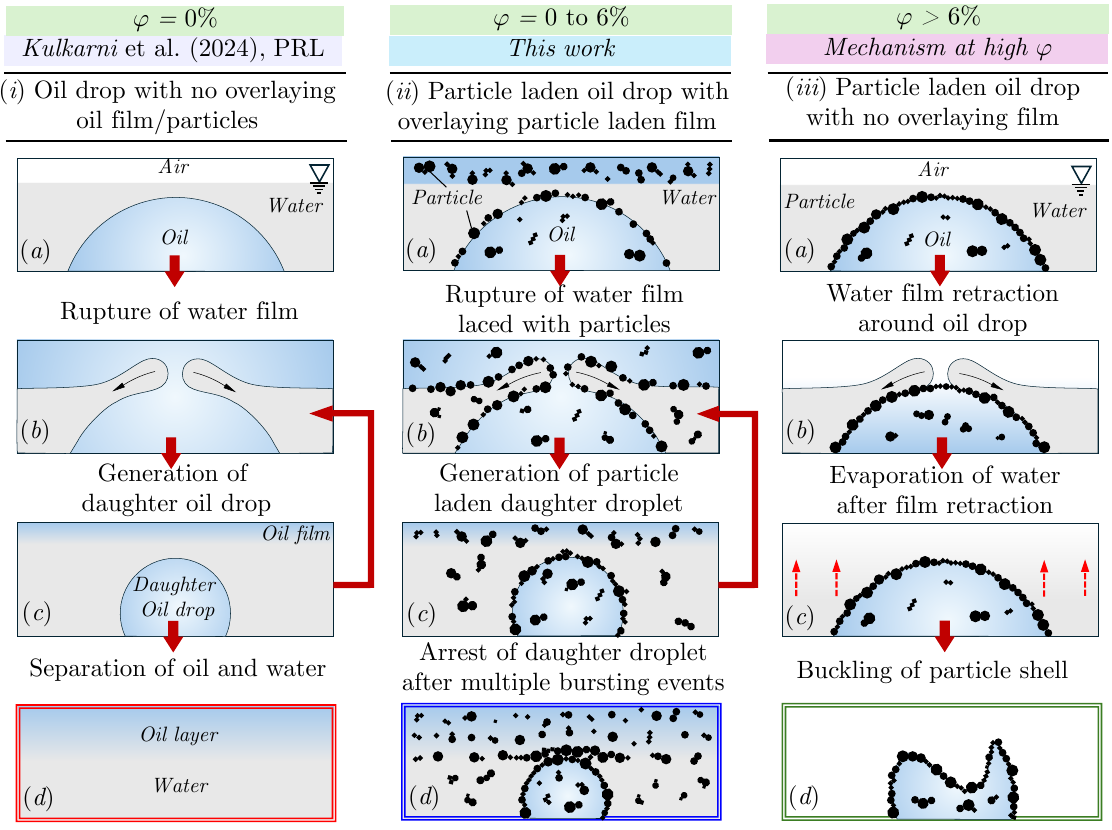}
   \vspace{-10pt}
  \caption{Schematic representations of three cases (\textit{i}) Oil drop with and overlaying oil film without any particles, Kulkarni et al, 2024 \citep{Kulkarni2024} (\textit{a}) Initial position of the oil drop as it rests near the water-air interface with the water slowly draining between the air and oil drop (\textit{b}) Rupture of the water film followed by its retraction and generation of capillary waves (\textit{c}) Separation of oil film from parent oil drop which repeats itself (as marked by maroon arrows) (\textit{d}) Finally leading to entire separation of oil from water as a film floating over it. (\textit{ii}) Particle laden oil drop with a overlaying particle laden oil film, THIS STUDY (\textit{a}) Initial position of the oil drop laden with particles in the oil film above and surface coverage less than 0.90 to prevent tight packing (\textit{b}) Rupture of water film bounded by particles on water-oil interface (\textit{c}) Formation of two separate entities from the parent particles infused drop – an oil infused film and an particle infused oil drop (\textit{d}) Final stage and arrest of drop due to particle laden oil  film on top and tight packing around the oil drop (\textit{iii}) Particle laden oil drop with no overlaying oil film (\textit{a}) Tight packing which disallows draining of the oil from the drop (\textit{b}) Retraction of water film over the particle infused oil drop (\textit{c}) Evaporation (shown by dotted red arrows) of water surrounding particle laden drop (\textit{d}) Buckling of particle shell in the final stages. It is important to note that despite tight packing a stable oil-water emulsion is not guaranteed in this configuration. }
  \vspace{-5pt}
  \label{FigS0}
\end{figure}

\noindent $\bullet$ \textit{Cause of cessation of bursting \textup{:} Role of particle in the oil drop and the overlaying particle laden layer at \textup{0} $< \varphi <$ \textup{6} \textup{\%}} \vspace{5pt}
\\
\noindent As the bursting commences two distinct fluid entities are created – one is the oil drop itself and the other is an oil on top. The stage-wise transition as the parent oil drop approaches the water-air interface and forms a daughter droplet is depicted in Fig. \ref{FigS0}\textcolor{blue}{(\textit{i}) (\textit{a})-(\textit{d})}. This follows from our earlier findings on such systems as reported in Kulkarni et al. \citep{Kulkarni2024}. 

In our current work this is extended to water-oil drop systems where the parent oil drop is infused with hydrophilic clay particles in an attempt to emulate conditions in a typical anti-Bancroft emulsion. Figure \ref{FigS0} \textcolor{blue}{(\textit{ii})(\textit{a})-(\textit{d})} shows schematically the sequence of events followed once such a particle infused oil drops initially at rest (see Fig. \ref{FigS0}\textcolor{blue}{(\textit{ii})(\textit{a})}) is deformed by capillary waves. The rupture in this case follows the same dynamics as the one without particles (see Fig. \ref{FigS0} \textcolor{blue}{(\textit{i})(\textit{b})}) especially since the surface particle coverage is less than required for tight packing on the oil drop. This leads to the formation of a daughter droplet as shown in Fig. \ref{FigS0} \textcolor{blue}{(\textit{ii})(\textit{c})} with particles lacing its boundary and an oil film (with particles) above. The presence of particles in both the oil layer above and the daughter droplet is crucial to arrest to final daughter droplet formed after successive bursting events as shown in Fig. \ref{FigS0}\textcolor{blue}{(\textit{ii})(\textit{d})}. The significance of this lies in the fact that the particle laden oil layer above prevents further cascade by arresting any water film drainage and the tight packing on the oil drop prevents any oil drainage that may lead to subsequent size reduction. So both are equally important.
\vspace{5pt}
\\
\noindent $\bullet$ \textit{High particle weight fraction, $\varphi >$ \textup{6\%}}\vspace{5pt}
\\
\noindent For Pickering emulsions, a surface packing with a surface coverage of 0.9 or more is considered tight enough to prevent drainage of dispersed medium (in our case, oil) into the continuous medium (in our case water). This brings to light the possibility that the tight packing of the particles at the oil drop-water interface can arrest cascade and prevent production of further generations of daughter droplets. To understand what happens next, we refer to the steps outlined in Fig. \ref{FigS0}\textcolor{blue}{(\textit{iii})(\textit{a})-(\textit{d})}.  As water gradually drains above the particle infused oil drop at rest initially (see Fig. \ref{FigS0}\textcolor{blue}{(\textit{iii})(\textit{a})}), the water film thins ultimately rupturing (see Fig. \ref{FigS0}\textcolor{blue}{(\textit{iii})(\textit{b})})). At this stage the water level recedes owing to evaporation (see Fig. \ref{FigS0}\textcolor{blue}{(\textit{iii})(\textit{c})}) exposing the particle-infused surface to air. A particle shell finally emerges which loses it structural integrity with time and buckles under its own weight as depicted in (see Fig. \ref{FigS0}\textcolor{blue}{(\textit{iii})(\textit{d})}) as the oil it encompasses either seeps out or evaporates (albeit slower than water).

Therefore, in principle while it is true that further bursting is prevented at high particle volume fraction, $\varphi$, its formation is practically unproductive as it cannot form a stable oil-water mixture. This is so because, at high $\varphi$, initial surface coverage is high and the water surrounding the particle covered oil drop evaporates in the absence of a “protective” oil layer. We therefore restrict our attention to particle fractions which produce stable oil-water assemblies which roughly correspond to $\varphi < 6\%$.

\subsection{Effect of particle size distribution on scaling relation, $N_f \sim [\varphi^{-1/4}]$, Eq. (1) in manuscript}

For our chosen clay particles the following histogram is obtained with a mean of $\approx$15 $\mu$m (which is used in our theoretical scaling arguments) and standard deviation (std. dev.) of the distribution is 9.69 $\mu$m. The coefficient of variation (COV) $=$ std. dev./mean $=$ 9.69/15 is 0.646 which indicates the average particle size is a good representation of the overall data.
\begin{figure}[htp!]
      \vspace{-10pt}
   \centering
    \includegraphics[scale = 0.5]{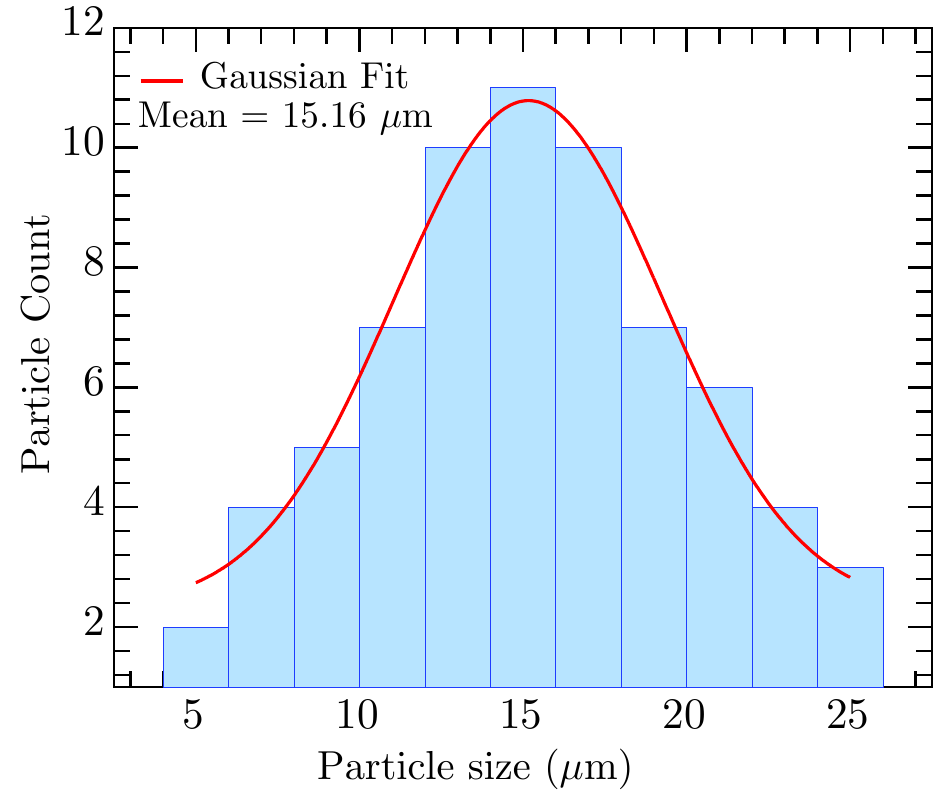}
   \vspace{-5pt}
  \caption{Particle size distribution for bentonite clay particles ($\approx$ 69 particles) as obtained from SEM image. Sample of the zoomed section is shown in Fig. 1(b) of the manuscript.}
  \vspace{-5pt}
  \label{FigS0H}
\end{figure}

We set the premise of our reasoning by quoting our manuscript, “larger $\varphi$ requires larger $R_f$ with a larger surface area, $\sim R_f^2$ to accommodate all the particles, therefore, it is reasonable to expect, $\varphi \sim n_{c,1} \sim n_{c,f} \sim R_f^2$ thereby simplifying our scaling relation to the form, $R_f^2 \sim \varphi$”.
To simplify this, we note that the total number of particles initially, $n_{c,1}$ is composed of individual particles of $n_{tot}$ (assumed without loss of generality) discrete sizes each $a_c(j)$ in size/diameter and number, constituting a particle size distribution of the form shown in Fig. \ref{FigS0H}. Each of these particles individually initially occupy a volume, $V_{c,1}^{(j)}$ and surface area, $S_{c,1}^{(j)}$ such that they constitute an individual volume fraction of $\varphi\left(j\right)$ given by, $\varphi^{(j)} \sim n_{c,1}^{(j)} V_{c,1}^{(j)}/\rho_o V_{\textrm{\textit{drop}}}$ where, $V_{\textrm{\textit{drop}}} = (4\pi/3) R_p^3$.

We can now apply the reasoning in the manuscript used for the entire set of particles to particles of a specific size, $a_c\left(j\right)$. Such a consideration means that volume fractions occupied by each of them initially and constituting a surface area, can be imagined to occupy a surface area, when the drop bursting is arrested strictly under the assumption that the initial particle size distribution remains the same. From this we obtain the scaling for each individual particle size as, which when summed over the entire range of particle sizes gives rise to,
\begin{equation}\label{Eq1ii}
\sum_{j=1}^{n_{tot}} \varphi_{c,f}^{(j)} \sim \sum_{j=1}^{n_{tot}} S_{c,f}^{(j)} \implies \varphi \sim S_{c,f} \sim R_f^2
\end{equation}
The rest of derivation remains the same as in our discussion in the paper which finally gives us the scaling relation, $N_f \sim [\varphi^{-1/4}]$ in Eq. (1) of the manuscript. 

\section{General considerations for film drainage in compound drops}
In this section we provide the details of derivation for arriving at scaling expressions Eqns 4 and 5 which are used to design the regime map in Fig. 4 of the main text.
\subsection{Drainage of liquid between two applied forces}
\begin{figure}[htp!]
   \centering
    \includegraphics[width = \textwidth]{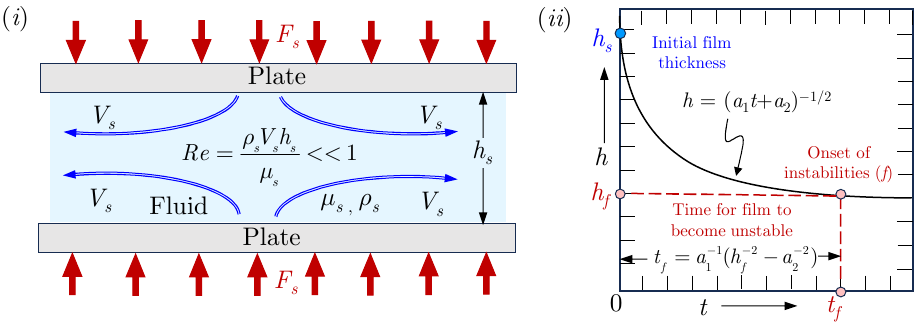}
   \vspace{-15pt}
\caption{(\textit{i}) Schematic of a typical squeeze/drainage flow problem with two forces equal in magnitude $F$, squeezing (\textit{s}) a fluid between of density, $\rho_{s}$ and dynamic viscosity, $\mu_{s}$ between two plates, $h_{s}$ apart (indicated by blue circles in (\textit{ii}) and (\textit{iii})), at a velocity, $V_s$ corresponding to $Re << 1$ and governed by Eq. \ref{eqn1} (\textit{ii}) Dependence of $dh/dt$ vs $t$ as given by Eq. \ref{eqn4} with $a_1 = 2F/6\mu_{s}R_{s}^4$, $a_2 = h_{s}^{-1}$. The time, $t_f$, indicated by a red circle is one at which instabilities are expected to begin manifesting themselves and is the final (\textit{f}) instant of time beyond which a simple description like the one explained here is not longer applicable (\textit{iii}) Dependence of $h$ vs $t$ as obtained in Eq. \ref{eqn5}.}
  \vspace{-5pt}
  \label{FigS1}
\end{figure}
\noindent All our mathematical formulations are based on drainage flow of the bulk liquid thin film between the compound (parent) drop and the air/water interface and, the oil film between the water drop encapsulated within the (larger) oil drop. Such a situation is readily identifiable as a squeeze flow problem between in narrow gap of a given initial thickness before it is squeezed($s$), $h_{s}$ compressed by forces of magnitude $F_{s}$ (see Fig. \ref{FigS1}\textcolor{blue}{(\textit{i})}) imposing a pressure, $p\left(r, t\right)$. The lubrication equations related to low $Re << 1$ flow\citep{Leal2007} is given by,
\begin{equation}\label{eqn1}
\frac{dh}{dt} = \dfrac{1}{\mu_{s} r}\dfrac{\partial}{\partial r}\left(\dfrac{rh^3}{12}\dfrac{\partial p}{\partial r}\right)
\end{equation}
Assuming, \textit{h} is only a function of \textit{t} and not of \textit{r} we can integrate Eq. \ref{eqn1} with respect to \textit{r} noting that \textit{p} is applied on a circular area with radius $R_{s}$ to obtain,
\begin{equation}\label{eqn2}
p\left(r, t\right) = \dfrac{3 \mu_{s}}{h^3}\dfrac{dh}{dt}\left(r^2-R_{s}^2\right)
\end{equation}
Integrating Eq. \ref{eqn2} over a circular region of radius, $R_{s}$, we get,
\begin{equation}\label{eqn3}
F = \int_0^{2\pi} \int_0^{R_{s}} p\left(r, t\right) rdrd\theta = -\dfrac{3\pi}{2}\dfrac{\mu_{s} R_{s}^4}{h^3}\dfrac{dh}{dt}
\end{equation}
From Eq. \ref{eqn3} the rate at which liquid is squeezed out, $dh/dt$ can be expressed as,
\begin{equation}\label{eqn4}
\frac{dh}{dt}=-\frac{F_{s} h^{3}}{c \mu_{s} R_{0}^{4}}	
\end{equation}
The constant $c$ is added in Eq. \ref{eqn4} to account for different geometrical configurations like liquid squeezed between two plates or a sphere and a flat surface (see Section \ref{magForces}). To obtain the explicit dependence of $h$ with $t$ we can integrate Eq. \ref{eqn4} between the limits 0 to \textit{h} and 0 to \textit{t} to obtain,
\begin{equation}\label{eqn5}
h = \left(\dfrac{2F_{s}}{c\mu_{s} R_{s}^4}t + \dfrac{1}{h_{s}^2} \right)^{-1/2}
\end{equation}
Eq. \ref{eqn5}, suggests that it would require infinite time to achieve rupture \textit{i}.\textit{e}. $t \to \infty$ when $h \to 0$ which is physically impossible. In practice, at a certain critical final (\textit{f}) thickness, $h_f$, and time, $t_f$, the film becomes unstable and ruptures and is marked by red circles in the plot of Eq. \ref{eqn5} in Fig. \ref{FigS1}\textcolor{blue}{(\textit{ii})}. Therefore, the time, $t_f$ taken to reach $h_f$ can be calculated by rearranging terms in Eq. \ref{eqn5} after setting, $h = h_f$ as,
\begin{equation}\label{eqn6}
t_f = \dfrac{c\mu_{s}R_{s}^4}{2F_{s} h_{s}^2}\left(\dfrac{h_{s}^2}{h_f^2} - 1\right)
\end{equation}
In the above, it is convenient to drop the term $(h_{s}^2/h_f^2) - 1$ in further analysis if $h_f = 0.1 h_{s}$, which is the typical final film thickness prior to rupture in most film drainage scenarios \citep{Chatzigiannakis2021}. Further, we shall use Eqns \ref{eqn4} and \ref{eqn6} developed in this section with suitable substitutions for $F_{s}$, $\mu_{s}$, $R_{s}$ and $h_{s}$ to establish the regime boundaries, \textcolor{maroon}{(\textbf{A})}, \textcolor{maroon}{(\textbf{B})}, \textcolor{maroon}{(\textbf{C})}, \textcolor{maroon}{(\textbf{D})} and, \textcolor{maroon}{(\textbf{E})}.

\section{Relation between forces and geometric variables}
To accomplish our eventual goal we define, $V_{r} := V_w/V_o$ and  $\rho_{r}:= \rho_{w}/\rho_o$, which help bring subsequent expressions for the different forces in their simplified dimensionless form. In so doing, we consider the mean density, $\rho_m$ of the compound drop defined the ratio of the sum of the masses of the oil layer ($\rho_{o} V_{o}$) and the encapsulated water drop ($\rho_{w} V_{w}$) by the sum of their individual volumes, $V_w$ and $V_o$,
\begin{equation}\label{eqn16}
\rho_{m}:=\frac{\rho_{o} V_{o}+\rho_{w} V_{w}}{V_{\text{o}}+V_{w}} 
\end{equation}
By dividing the numerator and denominator of Eq. \ref{eqn16} by $\rho_{o} V_{o}$ we arrive at,
\begin{equation}\label{eqn16a}
\dfrac{\rho_{m}}{\rho_{o}} = \frac{1+\rho_{wo} V_{r}}{1+V_{r}} 
\end{equation}
Similarly, we can the express the parent volume, $V_p = V_w + V_o$ in dimensionless form (by dividing both sides by $V_o$) as,
\begin{equation}\label{eqn18}
	\dfrac{V_{p}}{V_o} = 1 + V_{r}
\end{equation}
In literature, instead of $V_{r}$ often the ratio of volumes ($V_t$) of encapsulated water($V_w$) drop and parent compound drop($V_p$) defined as, $V_t := V_w/V_p$ is used. It can be related to $V_{r}$ in the following manner.
\begin{equation}\label{eqn63}
V_{t}=\dfrac{V_{w}}{V_{w}+V_{\text {o}}}=\dfrac{\left(V_w/V_o\right)}{1+\left(V_w/V_o\right)} = \dfrac{V_{r}}{1+V_{r}} \quad \textrm{or equivalently,}  \quad V_{r}=\dfrac{V_{t}}{1-V_{t}} 
\end{equation}

\subsection{Magnitude of various forces}\label{magForces}
\begin{figure}[htp!]
   \centering
    \includegraphics[width = \textwidth]{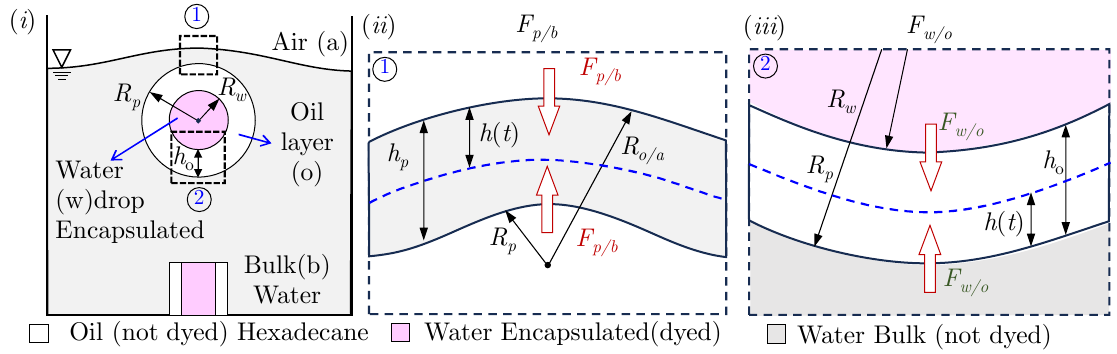}
   \vspace{-15pt}
  \caption{(\textit{i}) Schematic of the setup showing the compound drop of radius, $R_p$ with water drop of radius, $R_w$ encapsulated within an oil (hexadecane) drop such that an oil layer of thickness, $h_{\textrm{o}}$ is formed. (\textit{ii}) Magnified view of water (bulk liquid) thin film drainage between the compound drop and the air/water interface shown in (\textit{i}) as a dotted box \textcircled{\textcolor{blue}{1}} (\textit{iii}) Magnified view of of oil thin film drainage between the encapsulated water drop and the compound drop-bulk water interface shown in (\textit{i}) as a dotted box \textcircled{\textcolor{blue}{2}}.}
  \vspace{-5pt}
  \label{FigS2}
\end{figure}

The outcome of bursting of a parent compound drop (\textit{i}.\textit{e}. water drop encapsulated within the oil drop) after its release from a coaxial nozzle (shown in Fig. \ref{FigS2}\textcolor{blue}{(\textit{i})}) is decided by relative magnitude of the buoyant force of the parent compound drop (\textit{p}) acting on the thin film of bulk water, $F_{p/b}$ draining fluid residing between the bulk (\textit{b}) water(\textit{w})/air(\textit{a}) interface and, force acting on the thin film of oil contained between the encapsulated water (w) drop, $F_{w/o}$ and enveloping oil drop (o). In Fig. \ref{FigS3}\textcolor{blue}{(\textit{i})}-\textcolor{blue}{(\textit{iii})} we depict reduction of different scenarios to canonical situations which can be likened to that presented in Fig. \ref{FigS1}\textcolor{blue}{(\textit{i})}. The expressions for all scenarios are derived below and which shall be used in determining regime boundaries in Section \ref{RegBound}. 

\begin{enumerate}[leftmargin=*,align=left]
\item Expression for buoyant force, $F_{p/b}$
\\
The rise of the spherical compound drops of mean density, $\rho_m$ and volume, $V_p$ is a result of the imbalance of force due its own weight (downwards), $\rho_{m}V_{p}\mathrm{g}$ and buoyancy due to the displaced fluids (upwards), $\rho_{w}V_{p}\mathrm{g}$. The net upwards force so produced, $F_{p/b} = \left(\rho_{w}-\rho_{m}\right) V_{p} \mathrm{g}$ drains the bulk liquid (water) shown in Fig. \ref{FigS2}\textcolor{blue}{(\textit{ii})} as the magnified view of Fig. \ref{FigS2}\textcolor{blue}{(\textit{i})}, \textcircled{\textcolor{blue}{1}}. For, $R_{o/a} >> R_p$ the drainage of bulk liquid resembles the squeeze between a sphere and a flat surface as illustrated in Fig. \ref{FigS3}\textcolor{blue}{(\textit{i})}. Substituting for $\rho_m$ from Eq. \ref{eqn16a} and $V_p$ from Eq. \ref{eqn18} we can write the dimensionless buoyant force experienced by the compound drop as,
\begin{equation}\label{eqn6abs}
\dfrac{F_{p/b}}{\rho_{\text{o}} V_{\text{o}} \mathrm{g}} = \left(\rho_{r}-1\right)
\end{equation}
\item Expression for $F_{w/o}$
\\
\begin{figure}[htp!]
   \centering
    \includegraphics[width = \textwidth]{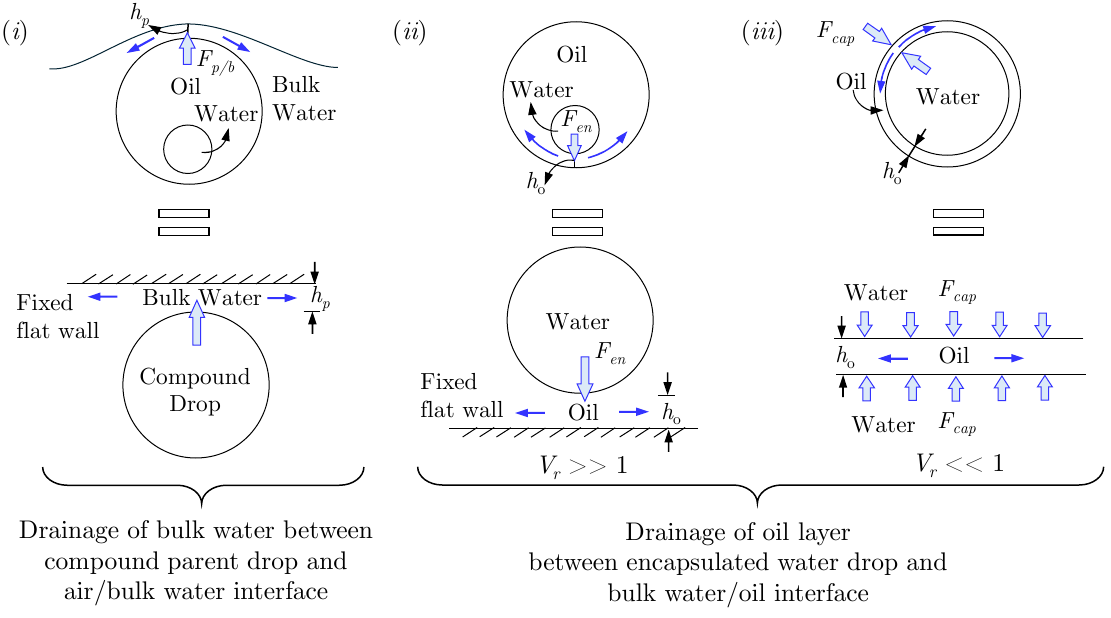}
   \vspace{-5pt}
  \caption{Simplified representations for (\textit{i}) bulk water thin film drainage as that of a rising sphere and a fixed flat wall since curvature of $R_{o/a} >> R_{p}$, refer Fig. \ref{FigS2}.  Oil thin film drainage at (\textit{ii}) $V_{r} >> 1$, when weight of the encapsulated water drop is the dominant force and $R_w << R_p$ (\textit{iii}) $V_{r} << 1$, when capillary force, $F_{\text{cap}}$ is dominant due to a very thin oil film, $h_{\text{o}} << 1$ and  $R_w \approx R_p$.}
  \vspace{-5pt}
  \label{FigS3}
\end{figure}
As the volume of the encapsulated water drop changes, the layer of oil encapsulating the water drop is reduced, consequently changing the experienced net force shown in Fig. \ref{FigS2}\textcolor{blue}{(\textit{ii})} as the magnified view of Fig. \ref{FigS2}\textcolor{blue}{(\textit{ii})}\textcircled{\textcolor{blue}{2}}. In this subsection, we resolve $F_{w/o}$ in the limits of high, $V_{r} >> 1$ as $F_{\text{en}}$ and low, $V_{r} << 1$ as $F_{\text{cap}}$
\begin{enumerate}[leftmargin=*,align=left]
\item When, $V_{r} << 1$, expression for $F_{\text{en}}$
\\
During the ascent of the compound parent drop, the encapsulated water drop, owing to it higher density continually sinks inside the compound parent drop imposing a force, $\left(\rho_{w}-\rho_{o}\right)V_w\mathrm{g}$ corresponding to its apparent weight (accounting for buoyancy) on the intervening layer of oil. Since the compound drop also moves upwards due to buoyancy, simultaneous to the sinking of encapsulated water drop, the reduction in the oil layer gap is hastened due to application of the corresponding force given by, $\rho_{w}V_p\mathrm{g}$. For $R_w << R_p$ as is the case for $V_{r} << 1$ the drainage flow here  resembles the squeezing of flow between sphere and a flat surface as illustrated in Fig. \ref{FigS3}\textcolor{blue}{(\textit{ii})}. The corresponding net force, $F_{\text{en}}$ imposed on this draining oil layer can be written as,
\begin{equation}\label{eqn6abw}
F_{\text{en}} = \left(\rho_{w}-\rho_{o}\right)V_w\mathrm{g} + \rho_{w}V_p\mathrm{g}
\end{equation}
Substituting for $V_p$ from Eq. \ref{eqn18} and definitions of $\rho_r$ and $V_{r}$ in replacing $\rho_w$ and $V_w$ respectively we can write,
\begin{equation}\label{eqn6a}
\dfrac{F_{\text{en}} }{\rho_{\text{o}} V_{\text{o}} \mathrm{g}} = \left(\rho_{r}-1\right) V_{r}+\rho_{r}\left(V_{r}+1\right)=2 \rho_{r} V_{r}-V_{r}+\rho_{r}
\end{equation}
\item When, $V_{r} >> 1$, expression for $F_{\text{cap}}$
\\ 
In contrast to low $V_{r}$ at high $V_{r}$ the oil layer thickness is sufficiently small for pressure due to capillarity to be high. Here, $R_w \approx R_p$ as is the case for $V_{r} >> 1$, the drainage of oil resembles the squeeze flow between two flat surfaces as illustrated in Fig. \ref{FigS3}\textcolor{blue}{(\textit{iii})}. The force, $F_{\text{cap}}$ generated due to capillarity then can be simply written as,
\begin{equation}\label{eqn6c}
F_{\text{cap}} = \sigma_{w/o}\left(2 \pi R_{p}\right)
\end{equation}
Here too, like in our previous calculations, we can make $F_{\text{cap}}$ dimensionless by diving by $\rho_{o} V_{o} \mathrm{g}$ to obtain,
\begin{equation}\label{eqn6d}
\dfrac{F_{\text{cap}}}{\rho_{o} V_{o} \mathrm{g}} = \dfrac{3}{2} \dfrac{\left(\rho_r -1\right) \left(1+ V_{r}\right)}{Bo_{w/o}}\quad\textrm{where,}\;\;Bo_{w/o} = \dfrac{\left(\rho_w - \rho_o\right)R_p^2\mathrm{g}}{\sigma_{w/o}}
\end{equation}
\end{enumerate}
\end{enumerate}
\subsection{Relation between $h_{\text {o}}, R_{w}$ and, $V_{r}$} \label{RwVr}
To make progress with algebra we need to express the relationships between the different geometric variables succinctly. In so doing, we note that $R_p = R_w + h_o$ and $V_{r}$ is expressed in the following manner.
\begin{equation}\label{eqn44a}
V_{r} := \dfrac{V_w}{V_o} = \dfrac{\frac{4}{3}\pi R_{w}^{3}}{\frac{4}{3}\pi \left(R_{p}^{3} - R_{w}^{3}\right)}
\end{equation}
Eq. \ref{eqn44a} can be further reduced using standard algebraic identities as,
\begin{equation}\label{eqn44b}
V_{r} =  \left[\dfrac{R_{p}^{3} - R_{w}^{3}}{R_{w}^{3}}\right]^{-1} = \left[\left(\frac{R_{p}-R_{w}}{R_{w}}\right)\left(\frac{R_{p}^{2}+R_{p} R_{w}+R_{w}^{2}}{R_{w}^{2}}\right)\right]^{-1}
\end{equation}
Since, $h_o = R_p - R_w$, Eq. \ref{eqn44b}, the above can be rewritten in the form,
\begin{equation}\label{eqn44c}
V_{r} =  \left[\left(\frac{h_{o} R_{p}^{2}}{R_{w}^{3}}\right)\left(1+\frac{R_{w}}{R_{p}}+\frac{R_{w}^{2}}{R_{p}^{2}}\right)\right]^{-1}
\end{equation}
\subsubsection{Simplification when $V_{r}<<1$}\label{Vrless1hoRw}
At low volume fractions two deductions can be made, $\dfrac{R_{w}}{R_{p}}<<1$ and $\dfrac{h_{o}}{R_{w}}>>1$. We use these in simplifying Eq. \ref{eqn44c} when $V_{r}<<1$.
\begin{equation}\label{eqn44}
V_{r} \approx \left[\frac{h_{o} R_{p}^{2}}{R_{w}^{3}}\right]^{-1} = \left[\left(\frac{h_{o}}{R_{w}}\right)\left(\frac{R_{p}}{R_{w}}\right)^{2}\right]^{-1}= \left[\left(\frac{h_{o}}{R_{w}}\right)\left(1+\frac{h_{o}}{R_{w}}\right)^{2}\right]^{-1}
\end{equation}
For, $\dfrac{h_{o}}{R_{w}} >> 1$, Eq. \ref{eqn44} can be reduced to the form,
\begin{equation}\label{eqn47}
\frac{h_{\text{o}}}{R_{w}} \approx V_{r}^{-\frac{1}{3}}
\end{equation}
\subsubsection{Simplification when $V_{r}>>1$}\label{VrhighhoRw}
At high volume fractions two deductions can be made, $R_w \approx R_p$ and $\dfrac{h_{o}}{R_{w}}<<1$. Similar to the previous section we use these to simplify Eq. \ref{eqn44c} when $V_{r}>>1$ to obtain the following,
\begin{equation}\label{eqn50}
\dfrac{h_{o}}{R_{w}} \approx \dfrac{1}{3V_{r}}
\end{equation}
\section {Criteria delimiting regimes of encapsulation, spreading and film bursting}\label{RegBound}
In this section we provide details on the scaling used to derive criteria for regime boundaries corresponding to Fig. 4 in the main text and shown by lines \textbf{\textcolor{maroon}{(A)}- \textcolor{maroon}{(E)}} in Fig. \ref{FigS4} below.
\begin{figure}[htpb!]
   \vspace{-5pt}
   \centering
    \includegraphics[width = \textwidth]{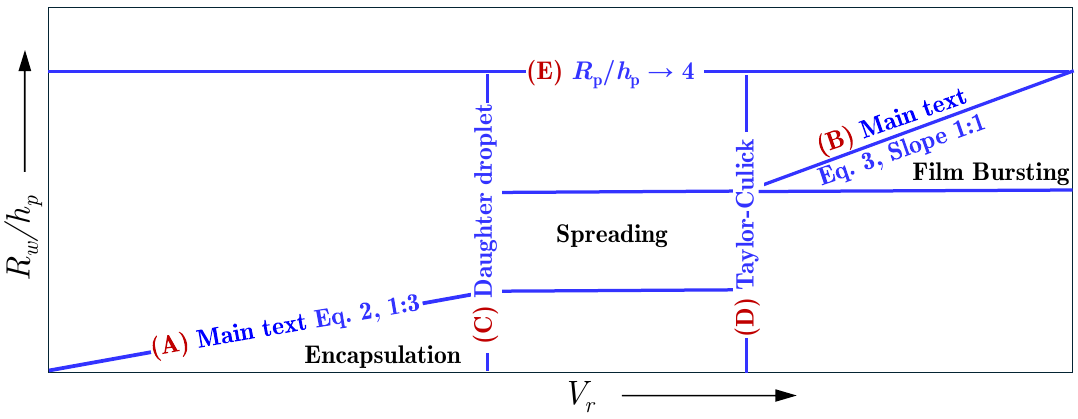}
   \vspace{-20pt}
  \caption{Regime boundaries demarcating the different regime transitions shown schematically, not to scale here and represented on log-log scale in Fig. 4 of the main text.\textbf{\textcolor{maroon}{(A)}} Boundary for continued encapsulation of water drop after bursting described here by Eq. \ref{eqn10kw}, Section \ref{sbVrless1} (refer Eq. 2 of main text) \textbf{\textcolor{maroon}{(B)}} Boundary for onset of oil film bursting in a compound drop described here by Eq. \ref{eqn39exc}, Section \ref{SectionB} (refer Eq. 3 of main text)\textbf{\textcolor{maroon}{(C)}} Horizontal boundary indicating cessation of encapsulation due to large daughter droplet size described here in Section \ref{SectionC}, Eq. \ref{eqn51a}, \textbf{\textcolor{maroon}{(D)}} Horizontal limit for onset of bursting due to film bursting given by Taylor-Culick velocity, details of which are provided in Section \ref{SectionD}, Eq. \ref{eqn51b} and, \textbf{\textcolor{maroon}{(E)}} Vertical limit for $R_w/h_p \to 4.4$ determined by the criterion, $Re_{b} \lesssim 1$, details of which are available in Section \ref{SectionE}, Eq. \ref{eqn21}.}
  \label{FigS4}
  \vspace{-5pt}
\end{figure}
\subsection{Derivation of scaling boundary, \textcolor{maroon}{(\textbf{A})}, $R_w/h_p  \sim V_r^{1/3}$}\label{sbVrless1}
The fate of a compound drop (\textit{i}.\textit{e}. water drop encapsulated within the oil drop) after it is released from the coaxial nozzle is decided by the time scale of drainage of oil film between encapsulated of water ($w$) drop immersed in the oil drop, $t_{w/\text{o}}$, which should be slower than the time scale of drainage of water (bulk) film between between the parent compound drop (\textit{p}) and the air/water interface, $t_{p/b}$. Mathematically, we write expressions for the two using Eq. \ref{eqn6} with appropriate substitutions for $F_{s}$, $\mu_{s}$, $R_{s}$ and $h_{s}$ and demand that at the boundary where daughter water droplet continues to be encapsulated within the parent oil drop after bursting, $t_{w/\text{o}} \sim t_{p/b}$ the following should be true,
\begin{equation}\label{eqn7}
\frac{c_{w/o} \mu_{o} R_{w}^{4}}{F_{w/\text{o}} h_{\text{o}}^{2}} \sim \frac{c_{p/b} \mu_{w} R_{p}^{4}}{F_{p/b} h_{p}^{2}}
\end{equation}
Or,
\begin{equation}\label{eqn8n}
\frac{F_{p /\text{b}}}{F_{w/\text{o}}} \sim \mu_{r} c_{s} \frac{h_{\text{o}}^{2}}{h_{p}^{2}}, \textrm{  where, }  c_s=\dfrac{c_{p/b}}{c_{w/o}} \textrm{  and, } \mu_r=\dfrac{\mu_w}{\mu_o}
\end{equation}
Before substituting for the forces, $F_{p/\text{b}}$ and $F_{w/\text{o}}$ above, where $c_s$ is the ratio of drainage/thinning/squeeze (\textit{s}) constants for the two configurations. We note that they can be applied only in the limit, $V_{r} << 1$ when $R_{w} << R_{p}$ and $h_{o}>>R_{w}$. In these conditions $F_{w/\text{o}} = F_{\textrm{en}}$ as derived in Eq. \ref{eqn6a}. At moderately high $V_{r}$, the expression for forces may be more complicated, containing dominant contributions from all, buoyancy, gravity and interfacial tension. However, at lower $V_{r}$ this is obviated by using Eqns. \ref{eqn6abs},\ref{eqn6a} and the assumption $c_s \approx 1$ to transform Eq. \ref{eqn8n} to,
\begin{equation}\label{eqn10}
\frac{\rho_r -1}{2 \rho_{r} V_{r}-V_{r}+\rho_{r}} \sim c_{s}\mu_{r} \frac{h_{\text{o}}^{2}}{h_{p}^{2}} = \mu_{r} \left(\frac{h_{\text{o}}^{2}}{R_{w}^{2}}\right) \left(\frac{R_{w}^{2}}{h_{p}^{2}}\right)
\end{equation}
Simplifying further using $V_r << 1$, we can replace ($2 \rho_{r} V_{r}-V_{r} + \rho_{r}$) by $\rho_{r}$ and substitute for $h_{\text{o}}/ R_w$ from Eq. \ref{eqn47} to arrive at the criterion below,
\begin{equation}\label{eqn10ab}
\dfrac{R_w}{h_p} \sim \left(\dfrac{\rho_r - 1}{\rho_r \mu_r^2}\right)^{1/2} V_r^{1/3}
\end{equation}
Since $\rho_r = 1.2$ and, $\mu_r = 3$ are constants for our test conditions for hexadecane and water, we can write,
\begin{equation}\label{eqn10kw}
\dfrac{R_w}{h_p} \sim V_r^{1/3}
\end{equation}
Eq. \ref{eqn10kw} corresponding to Fig. 4 and Eq. 2 in the main text is determined experimentally to be of the form, $R_w/h_p = 5V_r^{1/3}$ and shown by line \textbf{\textcolor{maroon}{(A)}} in Fig. \ref{FigS4} here.
\subsection{Derivation of scaling boundary, \textcolor{maroon}{(\textbf{B})}, $R_w/h_p \sim V_{r}$}\label{SectionB}
At high volume fractions, $V_{r}>>1$ with $R_{w} \approx R_{p}$ and $h_{o}<<R_{w}$ or $R_{p}$. We follow the procedure in Section \ref{sbVrless1} however with the crucial difference that the time scale, $t_{p/b}$ of drainage of the bulk water film between the parent compound drop (\textit{p}) air/water interface should be smaller than the time scale, $t_{\text{cap}}$ of drainage of the oil layer between the encapsulated water drop and water bath (bulk) due to capillary (cap) thinning. This means the bulk water film bursts earlier, triggering perturbations which can possibly lead to bursting of the oil film sandwiched between the water bath and encapsulated water drop. In mathematical terms, using Eq. \ref{eqn6} this can be written as,
\begin{equation}\label{eqn31}   
\frac{c_{w/\text{o}} \mu_{\text{o}} R_{w}^{4}}{F_{\text{cap}} h_{\text{o}}^{2}} \sim \frac{c_{p/b} \mu_{w} R_{p}^{4}}{F_{p/b} h_{p}^{2}}
\end{equation}
Or,
\begin{equation}\label{eqn32}
\frac{F_{p /\text{b}}}{F_{\text{cap}}} \sim \mu_{r} c_{s} \frac{h_{\text{o}}^{2}}{h_{p}^{2}}, \textrm{where, }  c_{s}:=\dfrac{c_{p/b}}{c_{w/\text{o}}}
\end{equation}
From Eqs. \ref{eqn6d}  and \ref{eqn6abs} we can substitute for $F_{p /\text{b}}$ and $F_{\text{cap}}$ and use the approximation for $c_{s}\approx 1$ to reduce Eq. \ref{eqn32} to, 
\begin{equation}\label{eqn34}
\dfrac{\left(\rho_{r}-1\right)}{\dfrac{3}{2} \dfrac{\left(\rho_r -1\right) \left(1+ V_{r}\right)}{Bo_{w/\text{o}}}} \sim \mu_{r} \dfrac{h_{\text{o}}^{2}}{h_{p}^{2}}
\end{equation}
Rewriting, $\dfrac{h_{\text{o}}^{2}}{h_{p}^{2}}$ on right hand side as, $\dfrac{h_{\text{o}}^{2}}{R_{w}^{2}}\dfrac{R_{\text{w}}^{2}}{h_{p}^{2}}$ we arrive at,
\begin{equation}\label{eqn34a}
\dfrac{Bo_{w/o}}{1+ V_{r}} \sim \dfrac{3\mu_{r}}{2} \dfrac{h_{\text{o}}^{2}}{R_{w}^{2}}\dfrac{R_{\text{w}}^{2}}{h_{p}^{2}}
\end{equation}
Substituting for $\dfrac{h_{o}}{R_{w}}$ from Section \ref{VrhighhoRw}, Eq. \ref{eqn50} we rewrite the above inequality as,
 \begin{equation}\label{eqn35}
\dfrac{Bo_{w/o}}{1+ V_{r}} \sim \dfrac{3\mu_{r}}{2} \left(\dfrac{1}{3V_{r}}\right)^2 \dfrac{R_{\text{w}}^{2}}{h_{p}^{2}}
\end{equation}
On rearranging the above in terms of $V_{r}$ we obtain the following,
\begin{equation}\label{eqn38}
\dfrac{V_{r}^{2}}{\left(1+V_{r}\right)} \sim \frac{\mu_{r}}{6 Bo_{w/o}}\left(\frac{R_{w}}{h_{p}}\right)
\end{equation}
Finally, we use the simplification, $V_{r}  >> 1$ such that, $\dfrac{V_{r}^{2}}{\left(1+V_{r}\right)} \approx V_{r}$, leading us to,
\begin{equation}\label{eqn39}
\dfrac{R_{w}}{h_{p}} \sim \left(\dfrac{6 Bo_{w/o}}{\mu_{r}}\right)V_{r}
\end{equation}
Like in the previous section $\mu_r = 3$ and $Bo_{w/o} \approx 0.45$  are constants for our test conditions for hexadecane and water, we can write,
\begin{equation}\label{eqn39exc}
\dfrac{R_{w}}{h_{p}} \sim V_{r}
\end{equation}
Eq. \ref{eqn39exc} corresponding to Fig. 4 and Eq. 3 in the main text is is determined experimentally to be of the form, $R_w/h_p = 0.02V_r + 2.0$ and shown as line \textbf{\textcolor{maroon}{(B)}} in Fig. \ref{FigS4}.
\subsection{Derivation of boundary \textcolor{maroon}{(\textbf{C})}, $V_r \approx 0.04$, Daughter droplet generation limit}\label{SectionC}
While the criterion given by inequality \ref{eqn10ab} sets the vertical limit for observing compound water encapsulated drops after bursting at the air/water interface, the limit on the horizontal axis of Fig. \ref{FigS4} corresponding to the $V_{r}$ at which this may cease to occur also needs to ascertained for the regime to be bounded completely. To do so, we use the dimensionless form of the scaling relation for the daughter droplet radius, $R_i/R_p$ produced after ``\textit{i}'' bursting events, $N_i$, \textit{i}.\textit{e}., $R_i/R_p \approx N_i^{-2}$ (also stated in the main text). Setting the value of $N_i = 2$, which corresponds to the first bursting event we obtain, $R_2 \approx R_p/4$. The encapsulated water drop of radius, $R_w$ must be less than the size of this daughter droplet size for perfect encapsulation to happen. Therefore we can write, $R_w/R_p < 1/4$. At low volume fractions, $V_{r} << 1$ from Eq. \ref{eqn47}, $\frac{h_{\text{o}}}{R_{w}} \approx V_{r}^{-\frac{1}{3}}$ derived in Section \ref{Vrless1hoRw} and noting, $R_p = R_w + h_o$ such that,  $R_p/R_w = 1 + h_o/R_w$, we can derive the horizontal limit, in terms of $V_r$, for continued water drop encapsulation in the parent compound drop after bursting as,
\begin{equation}\label{eqn51a}
V_{r} < 3^{-3} \approx 0.04 
\end{equation}
Eq. \ref{eqn51a} corresponding to Fig. 4 in the main text is shown as the vertical line \textbf{\textcolor{maroon}{(C)}} in Fig. \ref{FigS4}.

\subsection{Derivation of boundary \textcolor{maroon}{(\textbf{D})}, $V_r \approx 15$, Taylor-Culick velocity limit}\label{SectionD}
\begin{figure}[htpb!]
   \vspace{-5pt}
   \centering
    \includegraphics[width = \textwidth]{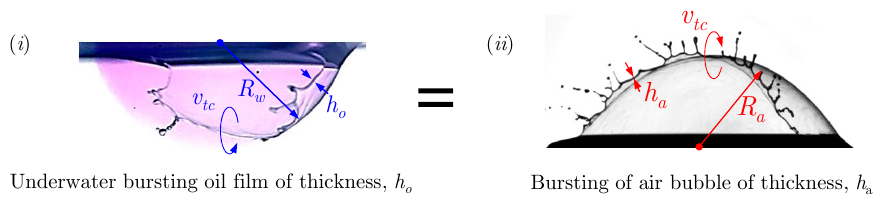}
   \vspace{-20pt}
  \caption{(\textit{i}) Retraction of hole in a bursting air bubble, Lhuissier and Villermaux \citep{Lhuissier2012}(reprinted with permission) (\textit{ii}) Retraction of a hole in a bursting underwater oil film.}
  \label{FigS5}
  \vspace{-5pt}
\end{figure}

At moderate (mod) $V_r$ the encapsulated drop radius, $R_w$ is large enough so that the oil film of thickness, $h_o$ is diminished to a size such that capillary effects dominate gravity and may be adequately described by the capillary(cap) time scale\citep{Deka2019}, $\tau_{mod} = \left(\rho_o R_p^3/\sigma_{w/o}\right)^{1/2}$. Herein, waves travel a distance of $1.5\pi R_p$, in time, $\tau_{mod}$ at an average velocity given by, $v_{mod}/2 = 1.5\pi R_p/\tau_{mod}$. Once $V_r$ increases to higher values, similar to a bursting air bubble a hole is nucleated in the  oil film (see Figs. \ref{FigS5} \textcolor{blue}{(\textit{i})} and \textcolor{blue}{(\textit{ii})}) which expands, retracting at the Taylor-Culick (\textit{tc}) velocity, $v_{tc} =\sqrt{2\sigma_{w/o}/\rho_oh_o}$. Setting, $v_{mod}/2 = v_{tc}$ we arrive at, $h_o/R_p = (9\pi^2/2)^{-1}$. For, $R_p \approx R_w$, from Eq. \ref{VrhighhoRw} in Section \ref{RwVr}, $V_r \approx (h_o/R_w)^{-1}/3$ giving us the criterion above which film bursting may be observed as,
\begin{equation}\label{eqn51b}
V_{r} >  \dfrac{3\pi^2}{2} \approx 15
\end{equation}
Eq. \ref{eqn51b} corresponding to Fig. 4 in the main text is shown as the vertical line \textbf{\textcolor{maroon}{(D)}} in Fig. \ref{FigS4}.
\vspace{-10pt}
\subsection{Derivation of boundary \textcolor{maroon}{(\textbf{E})}, $R_w/h_p \to 4.4$}\label{SectionE}
The maximum value attained by $R_w/h_p$ is when the encapsulated water drop occupies the entire volume of the compound drop, at $V_{r} >> 1$ equivalent to $R_w \to R_p$. Since the drainage flow equations mentioned in the preceding sections are only applicable after the compound drop rises to a certain distance close to the air/water interface it sets the upper limit for $R_p/h_p$ (and consequently, $R_w/h_p$) in our regime map. For any value above this, the bulk liquid film is expected to unconditionally stable. To ascertain this, we invoke the definition of the Reynolds number, $Re_{b}:=\rho_{w} v_{w} h_{p}/\mu_{w}$ and set it $\lesssim 1$ to guarantee dominant role of viscous forces. Using this condition we may derive the expression for $h_p$ as,
\begin{equation}\label{eqn21}
h_{p} \lesssim \frac{\mu_{w}Re_{b}}{\rho_{w} v_{w}} = \dfrac{0.001}{998 \times 0.002} \approx 0.5 \textrm{ mm}
\end{equation}
In the above we have used density of the bulk fluid made of water, $\rho_{w} \approx 998$ kg/m\textsuperscript{3}, $\mu_{w} \approx 1$ mPa-s and an approach velocity, $v_w \approx$  0.002 m/s (from our experiments). With $h_p \approx$ 0.5 mm and  $R_p \approx 2.2$ mm we get, $R_p/h_p \approx 4$. Since, $R_w \approx R_p$, we can write, $R_w/h_p \approx 4.4$, shown as the horizontal line \textbf{\textcolor{maroon}{(E)}} in Fig. \ref{FigS4} corresponding to Fig. 4 in the main text .
\vspace{-10pt}
\bibliographystyle{unsrtnat}
\bibliography{SI_APLRefs}